\documentclass[showpacs,aps,prc,preprint,nofootinbib,superscriptaddress]{revtex4-1}

\usepackage{amsmath}
\usepackage{amssymb}
\usepackage{slashed}
\usepackage{color}
\usepackage{changes}

\def\bm{\boldsymbol}
\newcommand{\bea}{\begin{eqnarray}}
\newcommand{\eea}{\end{eqnarray}}
\newcommand{\be}{\begin{eqnarray}}
\newcommand{\ee}{\end{eqnarray}}
\newcommand{\no}{\nonumber \\}

\newcommand{\del}{\partial}

\def\vp{{\bm p}}

\def\vk{{\bm k}}

\def\vr{{\bm r}}
\def\vs{{\bm\sigma}}
\def\la{\langle}
\def\ra{\rangle}

\begin{document}


\title { Time Reversal Invariance Violating and Parity Conserving effects
in Proton Deuteron  Scattering}


\author{Young-Ho Song}
\email[]{yhsong@ibs.re.kr}
\affiliation{Rare Isotope Science Project, Institute for Basic Science, Daejeon 305-811, Korea}

\author{Rimantas Lazauskas}
\email[]{rimantas.lazauskas@ires.in2p3.fr}
\affiliation{IPHC, IN2P3-CNRS/Universit\'e Louis Pasteur BP 28,
F-67037 Strasbourg Cedex 2, France}

\author{Vladimir Gudkov}
\email[]{gudkov@sc.edu}
\affiliation{Department of Physics and Astronomy, University of South Carolina, Columbia, SC, 29208}




\date{\today}

\begin{abstract}
Time reversal invariance violating parity conserving (TVPC)
effects are calculated for elastic proton deuteron scattering with
proton energies up to $2~$MeV.  Distorted Wave Born Approximation
is employed to estimate TVPC matrix elements, based on
hadronic wave functions, obtained by solving three-body
Faddeev-Merkuriev equations in configuration space
with realistic potentials.
\end{abstract}


\maketitle

\section{Introduction
\label{sec:Intro}}

The study of Time Reversal Invariance Violating and Parity Conserving (TVPC) effects is  important approach for a search
of new physics beyond the Standard Model.
In the Standard Model time reversal invariance violation
requires also parity violation. Therefore,
 an observation of TVPC effects can be interpreted as a direct signal
of new physics. TVPC effects in neutron-deuteron scattering have
been calculated recently~\cite{Song:2011jh}. In this paper we
consider similar effects of TVPC interaction in proton-deuteron
scattering which are related to $\vs_p\cdot[{\vp}\times{\bm
I}](\vp\cdot{\bm I})$ correlation
with tensor polarized target
, where  $\vs_p$ is the proton
spin, ${\bm I}$ is the target spin and $\vp$ is the proton
momentum. This correlation can be observed by measuring asymmetry
of protons polarized in parallel and anti-parallel to
$[{\vp}\times{\bm I}](\vp\cdot{\bm I})$ direction when transmitted through a
deuteron target. This is the simplest system to realize
aforementioned correlation related with TVPC effects for proton
scattering. Five-fold correlation $\vs_p\cdot[{\vp}\times{\bm
I}](\vp\cdot{\bm I})$ is equal to zero, unless the target spin $I$
 is larger or equal to one. As a consequence, this correlation
cannot be observed in nucleon-nucleon scattering.

 TVPC effects in proton deuteron  forward scattering for few hundreds $MeV$ proton energy range have been calculated~\cite{Beyer:1993zw,Uzikov:2015aua},
 in the relation to proposed experiment at COSY facility~\cite{pdEx:2013}.
We consider TVPC effects for proton energy range up to $2~MeV$
which could be calculated accurately in a formally exact framework
based on Faddeev-Merkuriev equations~\cite{Merkur:1980}
with realistic potentials.
This gives an opportunity to compare directly these TVPC effects with
the case of TVPC~\cite{Song:2011jh} effects in neutron-deuteron
scattering, as well as to the cases of parity violation in
proton-deuteron and neutron-deuteron~\cite{Song:2010sz}
scattering.

\section{Observables}
\label{sec:observables}

 In a contrast to neutron-deuteron scattering, the proton-deuteron scattering amplitude $f_{full}$
 diverges at zero scattering angle due to the Coulomb interaction.
 To avoid this divergence
 in the calculations of TVPC effects  we estimate an ``nuclear" amplitude $f=f_{full}-f_{Coul}$ with
    Coulomb amplitude being subtracted. Since Coulomb interaction does not violate time reversal invariance
    it cannot contribute to TVPC effects.
   For further calculations  we fix the direction of the proton
   momentum as axis $z$, and the direction of $[{\vp}\times{\bm I}](\vp\cdot{\bm I})$,
   as axis  $y$. Then, zero-angle scattering amplitudes
   $f_{\pm}(E,\theta=0)$, for protons,  polarized along and opposite
   to the direction of $[{\vp}\times{\bm I}](\vp\cdot{\bm I})$,  and propagating
   through the tensor polarized deuteron target are defined as \bea
   \label{eq:amp} f_{\pm}(E,\theta=0) \equiv \frac{1}{2} \sum'_{m_d}
            f\left(p\hat{z},(\frac{1}{2} \frac{\pm 1}{2})^{\hat{y}},(1 m_d)^{\hat{x}\hat{z}}\leftarrow
            p\hat{z},(\frac{1}{2} \frac{\pm 1}{2})^{\hat{y}},(1 m_d)^{\hat{x}\hat{z}}
            \right).
   \eea
   Here, $\sum'$ means that the state with $m_d=0$ is excluded from the summation,
   and the factor $\frac{1}{2}$ in the front of the summation is a  deuteron spin statistical factor.
   Then, using optical theorem~\cite{OptTheorem65},
    the  asymmetry in the transmission of polarized proton through tensor-polarized deuteron target can be written  as
\bea \label{eq:P}
P(E)
=\frac{\sigma^{nuc}_{+}-\sigma^{nuc}_{-}}{\sigma^{nuc}_{+}+\sigma^{nuc}_{-}}
=\frac{{\rm Im}[ f_{+}(E,\theta=0)-f_{-}(E,\theta=0)]}
      {{\rm Im}[ f_{+}(E,\theta=0)+f_{-}(E,\theta=0)]}.
\eea

The corresponding  ``nuclear" S-matrix (with subtracted Coulomb scattering part) is defined from the asymptotic form of
scattering wave function for  partial waves $\alpha'$ and $\alpha$,
where $\alpha=(L,S,J,T)$,
\bea
\frac{w_{\alpha',\alpha}(r;p)}{r}
&\rightarrow & \frac{1}{2}[\delta_{\alpha',\alpha}H_{l'}^{(-)}(\eta,\rho)
           +S_{\alpha',\alpha} H_{l'}^{(+)}(\eta,\rho)], \mbox{ for } {r\to \infty}
\eea
with
\bea
H_{l}^{(\pm)}(\eta,\rho)=\frac{1}{\rho}[F_{l}(\eta,\rho)\mp i G_l(\eta,\rho)],
\eea
where $F_l(\eta,\rho)$ and $G_l(\eta,\rho)$ are regular and irregular Coulomb functions
, $\eta=\frac{Z_1 Z_2\mu\alpha}{p}$ is a Sommerfeld parameter,
$\mu$ is a reduced mass, and $\rho= p r$.
Then, the ``nuclear" scattering amplitudes in Eq.(\ref{eq:amp}) are related
with ``nuclear" S-matrix as
\bea
f\left(\vp',1 m'_d, \frac{1}{2} m' \leftarrow \vp,1 m_d, \frac{1}{2} m\right)
&=&\sum_{LS,L'S',J} f_{L'S',LS}^J(p)
 \left( Z^{(J),L'S' m'_d m'}_{LSm_d m}(\hat{p}',\hat{p})\right)
\eea
where,
\bea
f_{L'S',LS}^J(p)&=&
\left(
  e^{i\sigma_{L'}}\frac{S^J_{L' S',L S}-\delta_{LL'}\delta_{S S'}}
  {2 i p} e^{i\sigma_L}
  \right),\no
Z^{(J),L'S' m'_d m'}_{LSm_d m}(\hat{p}',\hat{p})
&=& \sum_{L_z,L'_z,J_z}
    4\pi i^{-L'+L} Y_{L'L'_z}(\hat{p}')Y^*_{LL_z}(\hat{p})
      \no & &\times
    \la L L_z, S m_d+m|J J_z\ra
    \la 1 m_d, \frac{1}{2} m| S m_d+m\ra
     \no & &\times
    \la L' L'_z , S' m'_d+m'|J J_z\ra
    \la 1 m'_d, \frac{1}{2} m'| S' m'_d+m'\ra
\eea and $\sigma_l(\eta)\equiv \mbox{arg} \Gamma(l+1+i\eta)$ is a
Coulomb phase shift. Since the TVPC interaction is considered to
be weak, we can use Distorted Wave Born Approximation (DWBA) to
express the symmetry violating  scattering amplitudes related to
TVPC potential \bea
f^{\slashed{T}P}_{\alpha\beta}(k)=e^{i\sigma_\alpha}
        \left(\frac{\hat{S}_{\slashed{T}P}-1}{2i k} \right)_{\alpha,\beta}e^{i\sigma_\beta}
        \simeq
          -2\mu
          e^{i\sigma_\alpha}
          \left \la \psi^{(-)}_\alpha \Big|V_{\slashed{T}P}\Big|\psi^{(+)}_\beta\right\ra
          e^{i\sigma_\beta},
\eea
where
$\la \vr|\psi_\alpha^{(\pm)}\ra
=\sum_{\alpha'}\frac{w^{(\pm)}_{\alpha',\alpha}(r;p)}{r}{\cal Y}_{\alpha'}(\hat{r})
$ represents wave function solutions
with outgoing and incoming boundary conditions in partial wave $\alpha$
with ${\cal Y}_{\alpha'}(\hat{r})$
representing tensor spherical harmonics
in partial wave $\alpha'$.
Thus, by calculating matrix elements
$\left \la \psi^{(-)}_\alpha \Big|V_{wk}\Big|\psi^{(+)}_\beta\right\ra$,
we can obtain the ``nuclear" asymmetry $P$  of TVPC interaction in Eq.(\ref{eq:P}).

\section{Time reversal violating Parity Conserving potential
\label{sec:TVPCpot}}

The most general form of time reversal violating
and parity conserving part of nucleon-nucleon Hamiltonian
in the first order of relative nucleon momentum
can be written as \cite{Pherzeg66},
\bea
\label{eq:TVPC:general}
H^{\slashed{T}P}
&=&\left( g_1(r)+g_2(r)\tau_1\cdot\tau_2+g_3(r)T_{12}^z+g_4(r)\tau_{+}\right)
                 \hat{r}\cdot {\bar\vp}
   \no & &
   +\left(g_5(r)+g_6(r)\tau_1\cdot\tau_2+g_7(r)T_{12}^z+g_8(r)\tau_{+}\right)
                          \vs_1\cdot\vs_2\hat{r}\cdot \bar\vp
   \no & &
   +\left(g_9(r)+g_{10}(r)\tau_1\cdot\tau_2
        +g_{11}(r)T_{12}^z+ g_{12}(r)\tau_{+}\right)
    \no & &\quad \times
    \left(\hat{r}\cdot\vs_1{\bar\vp}\cdot\vs_2
         +\hat{r}\cdot\vs_2{\bar\vp}\cdot\vs_1
         -\frac{2}{3}\hat{r}\cdot{\bar\vp}\vs_1\cdot\vs_2
    \right)
    \no & &
   +(g_{13}(r)+g_{14}(r)\tau_1\cdot\tau_2
     +g_{15}(r)T_{12}^z+g_{16}(r)\tau_{+})
     \no & &\quad \times
    \left( \hat{r}\cdot\vs_1\hat{r}\cdot\vs_2
           \hat{r}\cdot{\bar\vp}
    -\frac{1}{5}(\hat{r}\cdot{\bar\vp}\vs_1\cdot\vs_2
                +\hat{r}\cdot\vs_1{\bar{\vp}}\cdot\vs_2
                +\hat{r}\cdot\vs_2\bar{\vp}\cdot\vs_1)
    \right)
    \no & &
   +g_{17}(r)
    \tau_{-}\hat{r}\cdot (\vs_\times\times{\bar\vp}
   +g_{18}(r)\tau_\times^z
    \hat{r}\cdot (\vs_{-}\times {\bar\vp}),
\eea
where the exact form of
$g_i(r)$ depends on the details of a particular theory of TVPC.

One should note, that pions, being spin zero
 particles, do not contribute to $TVPC$ on-shell interaction~\cite{Simonius:1975ve}.
Therefore to describe TVPC nucleon-nucleon interaction in meson
exchange potential model, by assuming CPT conservation, one should
consider contribution from heavier mesons: $\rho(770)$,
$I^G(J^{PC})=1^+(1^{--})$ and $h_1(1170)$,
$I^G(J^{PC})=0^-(1^{+-})$ (see, for
example~\cite{Haxton:1993dt,Haxton:1994bq,Gudkov:1991qc} and
references therein). The Lagrangian for strong and TVPC
interaction with explicit $\rho$ and $h_1$ meson exchanges is
expressed as \bea {\cal L}^{st}&=&-g_{\rho}
  {\bar N}(\gamma_\mu \rho^{\mu,a}-\frac{\kappa_V}{2M}\sigma_{\mu\nu}
  \del^\nu \rho^{\mu,a})\tau^a N
  -g_{h}{\bar N}\gamma^\mu\gamma_5 h_\mu N,
\eea
\bea
{\cal L}^{\slashed{T}P}&=&-\frac{\bar{g}_\rho}{2m_N}
        {\bar N}\sigma^{\mu\nu}\epsilon^{3ab}\tau^a \del_\nu \rho^b_{\mu} N
        +i\frac{\bar{g}_h}{2m_N}
       {\bar N}\sigma^{\mu\nu}\gamma_5\del_\nu
       h_\mu N,
\eea
where we neglected terms
${\bar N}\gamma_5\del^\mu h_\mu N$, which are small at low energy.
The parameters  $g$ and $\bar{g}$ are meson nucleon coupling constants for strong and TVPC interactions,
respectively.
 Then, one can separate TVPC potential due to $\rho$ and $h_1$ meson exchange as
\bea
V^{\slashed{T}P}_\rho&=&\frac{g_\rho{\bar g}_\rho m_\rho^2}{8\pi m_N} Y_1(m_\rho r)
                \tau_{\times}^z \hat{r}\cdot
                (\vs_{-}\times\frac{\bar\vp}{m_N}),\no
V^{\slashed{T}P}_{h_1}&=&-\frac{g_h\bar{g}_h m_h^2}{2\pi m_N}Y_1(m_h r)
                  (\vs_1\cdot\frac{\bar\vp}{m_N}\vs_2\cdot\hat{r}
                  +\vs_2\cdot\frac{\bar\vp}{m_N}\vs_1\cdot\hat{r}),
\eea
where $Y_1(x)=(1+\frac{1}{x})\frac{e^-x}{x}$,  $x_a=m_a r$.
Comparing  these potentials with eq. (\ref{eq:TVPC:general}), one can see that in the meson exchange  (ME) model, all
$g_i(r)^{ME}=0$, except for
\bea \label{eq:Yukawa}
g_{5}^{ME}(r)&=& \left(-\frac{ g_{h}\bar{g}_h m_h^2}{3m_N^2 \pi}\right)Y_1(m_h r)
              =c_5^h Y_1(m_h r)
,\no
g_{9}^{ME}(r)&=&\left(-\frac{g_{h}\bar{g}_h m_h^2}{2 m_N^2 \pi}\right)Y_1(m_h r)
              =c_9^h Y_1(m_h r),\no
g_{18}^{ME}(r)&=&\left(\frac{g_{\rho}\bar{g}_\rho m_\rho^2}{8 m_N^2 \pi}\right)Y_1(m_\rho r)
             =c_{18}^\rho Y_1(m_\rho r) .
\eea The possible contributions from heavier  vector iso-vector
mesons like $a_1$ and $b_1$ correspond to $g_{6}$ and $g_{10}$
functions of TVPC potential. However, for the sake of simplicity,
in this work we  focus only on the contribution from the exchange
of the lightest mesons, by considering $\rho$ and $h_1$.

Because the function $Y_1(\mu r)$ for $\rho$ and $h_1$ mesons is
singular at short ranges, the calculation of potential matrix
elements requires careful treatment at short distances. One way to
regulate the singular behavior of $Y_1(\mu r)$ Yukawa function is
by introducing regulated Yukawa function $Y_{1\Lambda}(r,m)$ with
a momentum cutoff $\Lambda$ as \bea \label{eq:Cutoff}
Y_{1\Lambda}(r,m)= -\frac{1}{m}\frac{d}{dr}
      \int \frac{d^3 k}{(2\pi)^3} e^{i\vk\cdot\vr} e^{-\frac{k^2}{\Lambda^2}}
      \frac{1}{k^2+m^2}.
\eea
From the point of view of effective field theory, we may regard
eq.(\ref{eq:TVPC:general}) as a leading order potential of EFT.
In this approach, cutoff represents our ignorance on short
distance dynamics and the low energy constants have to be
renormalized to absorb the cutoff dependence so that the final
results should not be sensitive to short distance uncertainty.
This approach, which  was adopted in our previous work on
neutron-deuteron scattering, is preferable from theoretical point
of view. However,  it introduces many unknown low energy constants
which have to be fixed from experiments. Therefore, to be able to
make a prediction for the value of TVPC observable, instead of
following a rigorous EFT approach, we use meson exchange model.
Then, calculating
 the potential matrix elements  using both
$Y_{1}(\mu r)$ and $Y_{1\Lambda}(r,\mu)$ with $\Lambda=1.5$ GeV,
 the difference of  two calculations
 can be attributed to the
uncertainty related with the short-range interactions.

\section{Results and discussions
\label{sec:results:tvpc}}
For the calculation of TVPC amplitudes in DWBA approach, we used
the non-perturbed (time reversal invariance conserving) 3-body wave functions for proton-deuteron scattering
 obtained by solving Faddeev-Merkuriev equations in configuration space~\cite{Merkur:1980} for $AV18$ nucleon-nucleon potential
in conjunction with $UIX$ three-nucleon force. The detailed
procedure for these calculations is described in  our
papers~\cite{These_Rimas_03,Song:2010sz,Song:2011sw}.

The main results of the calculations are summarized in table \ref{tbl:tv} where the imaginary part of time-reversal invariant
scattering amplitudes $(f_{+}+f_{-})(E,\theta=0)$ and
the TVPC scattering amplitudes $(f_{+}-f_{-})(E,\theta=0)$
in meson exchange models  are calculated with AV18 UIX potential.
To compare TVPC effects in proton-deuteron scattering with the case of neutron-deuteron scattering, we include the corresponding TVPC scattering amplitudes of
neutron-deuteron scattering at $E_{cm}=100$ KeV in the last line of table \ref{tbl:tv}. Note that there is a convention
difference with \cite{Song:2011jh}, and unpolarized  total proton-deuteron cross section
can be written as $\sigma_{tot}^{el}=\frac{1}{2}\frac{4\pi}{p}{\rm Im}(f_{+}+f_{-})(E,\theta=0)$.

To test how TVPC amplitudes depend on the choice of strong interaction potential we calculated them with three different  phenomenological
potentials AV18, AV18UIX and INOY.
We found  that the  time-reversal conserving scattering amplitudes calculated with these  three different
potentials  are in very good agreement
 for considered proton energy range $E_{cm}\leq 2$ MeV.
For example,
the amplitudes at $E_{cm}=1$ MeV (see second column of  table \ref{tbl:tv:potential}) shows that
AV18UIX and INOY potential results agrees well
with each other and comparison with AV18
implies that 3-body force effects contribute only at the level of
 $2 \%$. This result is not surprising because these amplitudes, which reproduce the total cross sections,
  are mostly sensitive to the long-range part of the interaction.

For the  TVPC and PV matrix elements, which are more sensitive to
the short range behavior of the potential, we can expect stronger
dependence on the strong interaction input. Moreover singularities
of Yukawa functions at short distances result in a finite residue
of the radial integrals for TVPC matrix elements at two-nucleon
contact which requires careful treatment of short range integrals
. Nevertheless, the results of calculations for the most
TVPC matrix elements with AV18 and AV18UIX potentials agree with
each other rather well. The operator $9$ (see table
\ref{tbl:tv:potential}) is an exception, which shows large
sensitivity to the presence of three nucleon force. Calculations
based on INOY NN interaction deviate from the AV18 case by 10\% -
20\%. It should be noted that similar discrepancies with INOY
potential
 was also observed in our previous calculations~\cite{Song:2010sz,Song:2011sw,Song:2011jh} of parity and time reversal violating effects
 in neutron deuteron interactions. This issue is
clearly related with a softness of INOY potential and the
qualitative difference of calculated nuclear wave functions at the
short distance.

In order to test the sensitivity of TVPC operators to short range
behavior of the potentials we calculated TVPC amplitudes with
Yukawa-type meson-exchange potentials Eq.(\ref{eq:Yukawa}) and
with regulated Yukawa potentials Eq.(\ref{eq:Cutoff}) with a cutoff
parameter $\Lambda = 1.5~GeV$. Thus,  comparing corresponding
results in tables (\ref{tbl:tv}) and (\ref{tbl:tv:cutoff}), one
can see rather good agreement between TVPC amplitudes calculated
with AV18UIX strong potential for different energies. The
comparison of tables (\ref{tbl:tv:potential}) and
(\ref{tbl:tv:potential2}) shows good agreement between the same
amplitudes calculated at $E_{cm}=1$ MeV with AV18UIX, AV18, and
INOY potentials. 

To be able  to test the consistency our calculations in the future when measurements of parity violating effects in proton deuteron scattering will be available,
 we calculated time reversal invariant  parity violating scattering amplitudes for opposite helicities $f^{pv}_{\pm}(E,\theta=0)$ defined as
\bea
f^{pv}_{\pm}(E,\theta=0)\equiv \frac{1}{3}\sum_{m_d}
 f\left(p\hat{z},(\frac{1}{2} \frac{\pm 1}{2})^{\hat{z}},(1 m_d)^{\hat{z}}\leftarrow
         p\hat{z},(\frac{1}{2} \frac{\pm 1}{2})^{\hat{z}},(1 m_d)^{\hat{z}}
         \right).
\eea
In this calculations we used a short range  iso-vector pion exchange part  of the DDH parity violating potential
\cite{Desplanques:1979hn}
\bea
V^{PV,DDH}_{1\pi}=\left( \frac{g_\pi h_\pi^1}{2\sqrt{2} m_N}\right)
                  (\tau_1\times\tau_2)^z(\vs_1+\vs_2)\cdot\hat{r}
                  \frac{d}{dr}\left(\frac{e^{-m_\pi r}}{4\pi r} \right).
\eea The results for ${\rm Im}
(f_{+}^{pv}-f_{-}^{pv})(E,\theta=0)$ are presented in table
\ref{tbl:pv:amp1}, where  the last line presents corresponding
parity violating amplitude for neutron-deuteron scattering at
$E_{cm}=100$ keV. One can see that PV amplitude is much less
sensitive to the particular choice of the strong interaction. This
is not surprising, since  PV effects are dominated by pion
exchange having much longer range.

Finally, by comparing our results  for proton deuteron and neutron deuteron scattering~\cite{Song:2011jh} at energy of $100~keV$ (see second and last rows in table \ref{tbl:tv}), one can see that corresponding amplitudes for these two processes have different sensitivity to TVPC $h_1$ an $\rho$ meson interactions. Therefore,
they   are rather complimentary to each other in the search for new physics, which can be manifested by TVPC interactions of $h_1$ an $\rho$ mesons with nucleons.

 \begin{table}
 \caption{\label{tbl:tv}
  Scattering amplitudes at various energies calculated with AV18UIX potential
  in $fm^{-1}$ units.
  The second column corresponds to time-reversal invariant
 ${\rm Im}(f_{+}+f_{-})(E,\theta=0)$  for tensor polarized deuteron target and
  other columns corresponds to Time-reversal violating scattering amplitudes
  $\frac{1}{c_{n}}{\rm Im}(f_{+}-f_{-})(E,\theta=0)$
  for operator $n$ and scalar function $Y_{1}(r,m)$.
  }
 \begin{ruledtabular}
 \begin{tabular}{rcccc}
 $E_{cm}$(keV)&${\rm Im}(f_{+}+f_{-})$ & n=5($m=m_h$) & n=9($m=m_h$)  & n=18($m=m_\rho$) \\
 \hline
$15  $&$       0.0907 $&$ 0.116\times 10^{-7} $&$ 0.131\times 10^{-6}  $&$-0.540\times 10^{-8} $\\
$100 $&$       1.76 $&$ 0.437\times 10^{-6} $&$0.348\times 10^{-5} $&$-0.136\times 10^{-6} $\\
$300 $&$       3.59 $&$ 0.177\times 10^{-5} $&$ 0.471\times 10^{-5} $&$-0.396\times 10^{-6} $\\
$1000$&$       6.75 $&$ 0.118\times 10^{-4} $&$ -0.658\times 10^{-5}  $&$0.482\times 10^{-5}  $\\
$2000$&$       8.04 $&$ 0.327\times 10^{-4} $&$-0.229\times 10^{-4} $&$ 0.296\times 10^{-5} $\\
\hline
nd $100$
& $2.85$ & $0.107\times 10^{-6}$ & $-0.217\times 10^{-5}$ & $-0.711\times 10^{-7}$
 \\
 \end{tabular}
 \end{ruledtabular}
 \end{table}

\newpage
 \begin{table}
 \caption{\label{tbl:tv:potential}
  Scattering amplitudes calculated at $E_{cm}=1$ MeV for various potential models
  in $fm^{-1}$ units.
  The second column corresponds to time-reversal invariant
   ${\rm Im}(f_{+}+f_{-})(E,\theta=0)$  for tensor polarized deuteron target and
    other columns correspond to TVPC scattering amplitudes
    $\frac{1}{c_{n}}{\rm Im}(f_{+}-f_{-})(E,\theta=0)$
    for operator $n$ and scalar function $Y_{1}(r,m)$.
  }
 \begin{ruledtabular}
 \begin{tabular}{rcccc}
 potential  & ${\rm Im}(f_{+}+f_{-})$ & n=5($m=m_h$) & n=9($m=m_h$)  & n=18($m=m_\rho$) \\
 \hline
 AV18UIX & $6.75$ & $0.118\times 10^{-4}$  & $-0.658\times 10^{-5}$ & $0.482\times 10^{-5}$ \\
 AV18    & $6.90$ & $0.102\times 10^{-4}$  & $0.258\times 10^{-5}$  & $0.403\times 10^{-5}$ \\
 INOY    & $6.75$ & $-0.324\times 10^{-5}$ & $0.482\times 10^{-4}$  & $0.103\times 10^{-4}$ \\
 \end{tabular}
 \end{ruledtabular}
 \end{table}

 \begin{table}
 \caption{\label{tbl:tv:cutoff}
  Scattering amplitudes at various energies calculated with AV18UIX potential
  in $fm^{-1}$ units.
  Each column corresponds to Time-reversal violating and parity conserving
  scattering amplitudes
  $\frac{1}{c_{n}}{\rm Im}(f_{+}-f_{-})(E,\theta=0)$
  for operator $n$ and scalar function $Y_{1\Lambda}(r,m)$ with $\Lambda=1.5$ GeV.
  }
 \begin{ruledtabular}
 \begin{tabular}{rccc}
 $E_{cm}$(keV)& n=5($m=m_h$) & n=9($m=m_h$)  & n=18($m=m_\rho$) \\
 \hline
$15  $& $ 0.174\times 10^{-7} $&$ 0.185\times 10^{-6}  $&$-0.540\times 10^{-8} $\\
$100 $& $ 0.633\times 10^{-6} $&$ 0.492\times 10^{-5} $&$-0.168\times 10^{-6} $\\
$300 $& $ 0.258\times 10^{-5} $&$ 0.680\times 10^{-5} $&$-0.246\times 10^{-6} $\\
$1000$& $ 0.173\times 10^{-4} $&$ -0.759\times 10^{-5}  $&$0.327\times 10^{-5}  $\\
$2000$& $ 0.484\times 10^{-4} $&$-0.274\times 10^{-4} $&$ 0.509\times 10^{-5} $\\
 \end{tabular}
 \end{ruledtabular}
 \end{table}

 \begin{table}
 \caption{\label{tbl:tv:potential2}
  Scattering amplitudes calculated at $E_{cm}=1$ MeV for various potential models
  in $fm^{-1}$ units.
  Each column corresponds to Time-reversal violating and parity conserving
  scattering amplitudes
  $\frac{1}{c_{n}}{\rm Im}(f_{+}-f_{-})(E,\theta=0)$
  for operator $n$ and scalar function $Y_{1\Lambda}(r,m)$ with $\Lambda=1.5$ GeV.
  }
 \begin{ruledtabular}
 \begin{tabular}{rccc}
 potential  &  n=5($m=m_h$) & n=9($m=m_h$)  & n=18($m=m_\rho$) \\
 \hline
 AV18UIX &  $0.173\times 10^{-4}$ & $-0.759\times 10^{-5}$ & $0.327\times 10^{-5}$ \\
 AV18    &  $0.150\times 10^{-4}$ & $0.242\times 10^{-5}$  & $0.243\times 10^{-5}$ \\
 INOY    &  $0.875\times 10^{-5}$ & $0.282\times 10^{-4}$  & $0.996\times 10^{-5}$ \\
 \end{tabular}
 \end{ruledtabular}
 \end{table}

  \begin{table}
 \caption{\label{tbl:pv:amp1}
  Parity violating scattering amplitudes
  $\frac{1}{c_{1}^{DDH} }{\rm Im}(f^{pv}_{+}-f^{pv}_{-})(E,\theta=0)$
  from PV DDH potential of iso-vector pion exchange in $fm^{-2}$ units, where
  $c_{1}^{DDH}=\frac{g_\pi h_\pi^1}{2\sqrt{2} m_N}$.
  }
 \begin{ruledtabular}
 \begin{tabular}{rccc}
 $E_{cm}$(keV) & AV18UIX & AV18  & INOY \\
 \hline
$15  $&$ 0.130\times 10^{-2} $&$           $&$           $\\
$100 $&$-0.425\times 10^{-1} $&$           $&$           $\\
$300 $&$-0.248\times 10^{+0} $&$           $&$           $\\
$1000$&$-0.729\times 10^{+0} $&$-0.728\times 10^{+0} $&$-0.751\times 10^{+0} $\\
$2000$&$-0.941\times 10^{+0} $&$           $&$           $\\
\hline
nd $100$ & $0.124\times 10^{-1}$ &  & \\
 \end{tabular}
 \end{ruledtabular}
 \end{table}

\begin{acknowledgments}
This material is based upon work supported by the U.S. Department
of Energy Office of Science, Office of Nuclear Physics program
under Award Number DE-FG02-09ER41621. The work of YS was supported
by the Rare Isotope Science Project of the Institute for Basic
Science funded by the Ministry of Science, ICT and the Future
Planning and National Research Foundation of Korea
(2013M7A1A1075764).
This  work  was  granted  access  to  the  HPC
resources  of TGCC and IDRIS under  the allocation
2015-x2015056006 made by GENCI. We thank the staff members of the
TGCC and IDRIS for their constant help.
\end{acknowledgments}

\bibliography{TViolation}

\end{document}